\documentclass[aps, showpacs,showkeys,pra] {revtex4}
\usepackage{amsmath }
\usepackage{amssymb}
\usepackage{epsfig}
\usepackage{natbib}
\newcommand*{\be}{\begin{equation}}
\newcommand*{\ee}{\end{equation}}

\begin{document}
\bibliographystyle{revtex}
\title{Improved variational approach for the Cornell potential}
\author{V.V. Kudryashov}
 \email{kudryash@dragon.bas-net.by}
 \author{V.I. Reshetnyak}
 \email{v.reshetnyak@dragon.bas-net.by}
\affiliation{Institute of Physics, National Academy of Sciences of
Belarus \\68 Nezavisimosti  Ave., 220072, Minsk,  Belarus }

\begin{abstract}{The approximate  radial wave functions for the Cornell potential describing quark-antiquark
 interaction are constructed in the framework of a variational method. The optimal values of
the  variational parameters are fixed by the fulfillment of the requirements of the virial theorem and
  the minimality condition for integral discrepancy. The results of calculation with the simple
  trial function are in a good agreement with exact numerical results.}
\end{abstract}

\maketitle

Many investigations have been devoted to the quantum mechanical
non-relativistic description of quark-antiquark bound states  with
the Cornell potential \cite{1} by means of different approximation
procedures (see, e.g., \cite{2,3,4,5} and references therein)
based on perturbation theory, the variational method and their
combinations. Recently a new improved variational approach
\cite{6} was proposed. This method was applied to the radial
Schr\"odinger equation with nonsingular power-law potentials $r^s$
with positive $s$ \cite{7}. In the present work, we apply our
approach to the Cornell potential with the Coulomb singularity.

 Following \cite{1} we study the radial
Schr\"odinger equation in the dimensionless form
\begin{equation}
\hat H \psi(r) = E \psi(r) , \quad
\hat H  = - \frac{d^2 }{d r^2}  + \frac{l (l+1)}{ r^2} + V(r)
 \end{equation}
for the Cornell (Coulomb plus linear) potential

\begin{equation}
V(r) =  - \frac{k}{r} + r .
 \end{equation}

 We use the simple functions \cite{8}
 \begin{equation}
  <r|\psi_{l,n}(a,b)> = \psi_{l,n}(a,b,r) = N \sqrt{a} G_{l,n}(b,a r)
\end{equation}
as the trial functions, where
\begin{equation}
 G_{l,n}(b,x)  = x^{l+1}\exp(-x^b)L_n^{\frac{2l+1}{b}}(2x^b).
 \end{equation}
 Here $L_n^{\frac{2l+1}{b}}(2 x^b)$ is the Laguerre polynomial, $a$ and $b$
are variational parameters, $N$ is a normalization factor
  $(<\psi_{l,n}(a,b)|\psi_{l,n}(a,b)>=1)$. Note that these functions reproduce the exact results for the Coulomb potential if we choose $b=1$.
In order to fix the parameter values we consider two conditions to which the trial functions
should satisfy.

The first is  the virial theorem. The virial condition
 \begin{equation}
  < \psi_{l,n}(a,b)| -\frac{d^2}{d r^2} +  \frac{ l
(l+1)}{r^2}- \frac{1}{2} \frac{k}{r} -\frac{1}{2} r|\psi_{l,n}(a,b)> =0
\end{equation}
leads to the following equation for $a$ :
 \begin{equation}
 a^3 = k u_2(b) a^2 + u_3(b)
 \end{equation}
 where
$$
u_2(b) = \frac{\int_0^{\infty}G_{l,n}^2(b,x) x^{-1} d x }
{2 \int_0^{\infty}G_{l,n}(b,x)(- \frac{d^2 }{d x^2} +\frac{l(l+1)}{x^2})G_{l,n}(b,x)  d x} ,
$$
$$
u_3(b) = \frac{\int_0^{\infty}G_{l,n}^2(b,x) x d x }
{2 \int_0^{\infty}G_{l,n}(b,x)(- \frac{d^2 }{d x^2} +\frac{l(l+1)}{x^2})G_{l,n}(b,x)  d x} .
$$
The exact solution of this equation  allows us to express the  parameter  $a$ via the parameter $b$ :
\begin{eqnarray}
a_0(b) = \frac{1}{3} k u_2(b) &+& \left( \frac{k^3 u^3_2(b)}{27} + \frac{u_3(b)}{2} +
\sqrt{\frac{k^3 u^3_2(b) u_3(b)}{27} + \frac{u^2_3(b)}{4}} \right)^{1/3} \nonumber \\
&+&\left( \frac{k^3 u^3_2(b)}{27} + \frac{u_3(b)}{2} -
\sqrt{\frac{k^3 u^3_2(b) u_3(b)}{27} + \frac{u^2_3(b)}{4}} \right)^{1/3}
\end{eqnarray}
 Thus,
the considered problem  is transformed into a one-parameter
problem.
Note that the virial condition is
equivalent to the  usual condition
 \begin{equation}
 \frac{\partial <\psi_{l,n}(a,b)| \hat
   H|\psi_{l,n}(a,b)>}{\partial a}{}=0 .
\end{equation}

Denote the one-parameter trial functions with $a = a_0(b)$ as
$$
 <r|\psi^0_{l,n}(b)> =  \psi^0_{l,n}(b,r) = <r|\psi_{l,n}(a_0(b),b)> = \psi_{l,n}(a_0(b),b,r) .
$$
Then the energy has the form
\begin{equation}
E^{(1)}_{l,n}(b) =  <\psi^0_{l,n}(b)| \hat  H|\psi^0_{l,n}(b)> .
\end{equation}
 In addition to the energy,  expectation
values of   the  squared Hamiltoninan characterizing the goodness
of the approximate eigenfunctions, can be calculated:
\begin{equation}
E_{l,n} ^{(2)}(b) = \sqrt{<\psi^0_{l,n}(b)|\hat H^2|\psi^0_{l,n}(b)>} .
\end{equation}

  As the second requirement imposed on the trial function,
we select the requirement that integral discrepancy
\begin{equation}
d_{l,n}(b) = \frac{<\psi^0_{l,n}(b)|\hat H^2|\psi^0_{l,n}(b)>}
{(<\psi^0_{l,n}(b)|\hat H|\psi^0_{l,n}(b)>)^2}-1,
\end{equation}
is minimum \cite{6}. The quantity $d_{l,n}(b)$  characterizes  goodness of the approximation and is equal to
zero for an exact solution of the Schr\"{o}dinger equation. The integral discrepancy is simply connected with  the local discrepancy \cite{9}
\begin{equation}
<r|D_{l,n}(b)> =   D_{l,n}(b,r)=  \frac{\hat
H\psi^0_{l,n}(b,r)}{{<\psi^0_{l,n}(b,r)|\hat H|\psi^0_{l,n}(b,r)>}} -
\psi^0_{l,n}(b,r)
\end{equation}
by relation
\begin{equation}
d_{l,n}(b) = < D_{l,n}(b)|D_{l,n}(b)> .
\end{equation}
 The corresponding equation for selection of parameter
 $b$  is
\begin{equation}
 \frac{d}{d b} d_{nl}(b)=0.
 \end{equation}
Note that we must find the absolute minimum of $d_{l,n}(b)$ corresponding to the minimal rotation of a trial
vector under the action of the Hamilt0nian in Hilbert space \cite{9}.

The minimality condition for integral discrepancy
is alternative to the minimal sensitivity condition \cite{10}
\begin{equation}
 \frac{d}{d b} E^{(1)}_{l,n}(b)=0.
\end{equation}
 There can  be several stationary  points in the case of the function $ E^{(1)}_{l,n}(b)$, and there is no method to select one of them. Besides,
energy  is not a preferred quantity in comparison
with the other expectation values. It is also well known, that
calculation of the energy with increased accuracy does not always
lead to the improvement of  other characteristics.

 We compare our results
  $E^{(1)}_{l,n}$ with the results $E^{vfm}_{l,n}$ of some variational method (VFM)
  \cite{2} and with  the results $E^{num}_{l,n}$ of numerical integration
  of the Schr\"odinger equation \cite{1}.  We
   also compare  our  values of the quantity
 \begin{equation}
   <v^2>_{l,n} = \int_0^{\infty}\left(\frac{d \psi^0_{l,n}(b,r)}{d r}\right)^2 d r
   \end{equation}
with  values from \cite{1}.

\begin{table}[t]
\begin{center}
\caption{ Linear potential $(k=0)$.}
\begin{tabular}{c| c| c| c| c| c| c| c}\hline
$l$ &$n$ &$E^{num}_{l,n}$ &$E^{(1)}_{l,n}$ &$E^{(2)}_{l,n}$ &$d_{l,n}$
&$<v^2>^{num}_{l,n}$ &$<v^2>_{l,n}$ \\
\hline
$0$&$0$&$2.3381$&$2.3383$&$2.3387$&$2.9465 \cdot 10^{-4}$&$0.7794$&$0.7794$ \\
$0$&$1$&$4.0879$&$4.0881$&$4.0890$&$4.4511 \cdot 10^{-4}$&$1.3626$&$1.3627$\\
\hline
$1$&$0$&$3.3613$&$3.3614$&$3.3615$&$5.5573 \cdot 10^{-5}$&$0.4921$&$0.4923$ \\
$1$&$1$&$4.8845$&$4.8846$&$4.8849$&$1.5180 \cdot 10^{-4}$&$1.1151$&$1.1057$ \\
\hline
$2$&$0$&$4.2482$&$4.2483$&$4.2483$&$1.8679 \cdot 10^{-5}$&$0.4089$&$0.4090$ \\
$2$&$1$&$5.6297$&$5.6298$&$5.6300$&$6.9838 \cdot 10^{-5}$&$1.0097$&$0.9997$ \\
\hline
\end{tabular}
\end{center}
\end{table}
\begin{table}[t]
\begin{center}
\caption{ Cornell potential.}
\begin{tabular}{c| c| c| c| c| c| c| c| c| c}\hline
$k$ &$l$ &$n$ &$E^{num}_{l,n}$ &$E^{VFM}_{l,n}$  &$E^{(1)}_{l,n}$ &$E^{(2)}_{l,n}$ &$d_{l,n}$
&$<v^2>^{num}_{l,n}$ &$<v^2>_{l,n}$ \\
\hline
$0.2$&$0$&$0$&$2.1673$&$2.1409$&$2.1673$&$2.1676$&$2.4358 \cdot 10^{-4}$&$0.8389$&$0.8389$ \\
$0.2$&$0$&$1$&$3.9702$&$3.9643$&$3.9704$&$3.9708$&$1.6310 \cdot 10^{-4}$&$1.4028$&$1.4032$ \\
\hline
$0.2$&$1$&$0$&$3.2582$&$3.1565$&$3.2580$&$3.2580$&$2.7539 \cdot 10^{-5}$&$0.5028$&$0.5031$ \\
$0.2$&$1$&$1$&$4.8019$&$4.7414$&$4.8019$&$4.8021$&$9.5487 \cdot 10^{-5}$&$1.1284$&$1.1212$ \\
\hline
$0.2$&$2$&$0$&$4.1703$&$4.0037$&$4.1703$&$4.1703$&$1.3279 \cdot 10^{-5}$&$0.4136$&$0.4137$ \\
$0.2$&$2$&$1$&$5.5634$&$5.4491$&$5.5634$&$5.5636$&$5.4213 \cdot 10^{-5}$&$1.0172$&$1.0086$ \\
\hline\hline
$1$&$0$&$0$&$1.3979$&$1.3460$&$1.4009$&$1.4173$&$2.3504 \cdot 10^{-2}$&$1.1716$&$1.1801$ \\
$1$&$0$&$1$&$3.4751$&$3.4765$&$3.4742$&$3.4841$&$5.7230 \cdot 10^{-3}$&$1.5870$&$1.5809$ \\
\hline\
$1$&$1$&$0$&$2.8255$&$2.7185$&$2.8257$&$2.8258$&$5.5640 \cdot 10^{-5}$&$0.5524$&$0.5520$ \\
$1$&$1$&$1$&$4.4619$&$4.3940$&$4.4619$&$4.4620$&$2.7702 \cdot 10^{-5}$&$1.1864$&$1.1868$ \\
\hline
$1$&$2$&$0$&$3.8506$&$3.6835$&$3.8506$&$3.8506$&$1.5440 \cdot 10^{-6}$&$0.4340$&$0.4341$ \\
$1$&$2$&$1$&$5.2930$&$5.1730$&$5.2930$&$5.2930$&$1.1421 \cdot 10^{-5}$&$1.0492$&$1.0456$ \\
\hline
\end{tabular}
\end{center}
\end{table}

 Table 1 shows that the proposed approximation gives fairly accurate results in the case of the purely linear potential. Table 2 demonstrates the efficiency of our approach in the case of the combined Coulomb plus linear potential.

 For example we compare our variant (14) and usual variant (15) in the case  $ k=1, l=1, n=1$:
 $$
 E^{(1)}_{1,1}(14)=4.4619,~  E^{(2)}_{1,1}(14)=4.4620,~ <v^2>_{1,1}(14)= 1.1868~ (a= 0.5439, b= 1.6769) ,
 $$
 $$
 E^{(1)}_{1,1}(15)=4.4583,~  E^{(2)}_{1,1}(15)=5.0658,~ <v^2>_{1,1}(15)= 0.9905~ (a=1.3352, b=0.9999)
 $$
 while the numerical solution \cite{1} gives $E^{num}_{1,1}= 4.4619$, $<v^2>^{num}_{1,1} =1.1864$ .

 We see that in spite of the simplicity of the used trial functions, the optimal choice of the variatonal parameters
 leads to the satisfactory approximation. At the same time we assume that the approximation can be improved by means of some complication of the trial function reproducing the behavior at the origin more accurately.

\
\end{document}